\documentstyle[multicol,prb,aps,epsf] {revtex}
\begin{document}
\bibliographystyle{prsty}
\draft
\title{Behavior of a frustrated quantum spin chain with bond dimerization}
\author{Tota Nakamura $^1$ and Satoshi Takada $^2$}
\address{$^1$ Department of Applied Physics, Tohoku University,
         Sendai, Miyagi 980-77, Japan\\
         $^2$ Institute of Physics, University of Tsukuba,
         Tsukuba, Ibaraki 305, Japan }
\date{\today}
\maketitle
\begin {abstract}
We clarified behavior of the excitation gap in a frustrated 
$S=1/2$ quantum spin chain 
with bond dimerization by using the numerical diagonalization of finite
systems and a variational approach.
The model interpolates between the independent dimer model
and the $S=1$ spin chain by changing a strength of the dimerization.
The energy gap is minimum at the fully-frustrated point,
where a localized kink and a freely mobile anti-kink
govern the low-lying excitations.
Away from the point,
a kink and an antikink form a bound state by an effective triangular
potential between them.
The consequential gap enhancement and 
the localization length of the bound state is obtained exactly
in the continuous limit.
The gap enhancement obeys a power law with exponent $2/3$. 
The method and the obtained results are common to other 
frustrated double spin-chain systems, 
such as the one-dimensional $J_1$-$J_2$ model, or the frustrated ladder model.
\end  {abstract}

\begin{multicols}{2}

\narrowtext

\section{Introduction}
Recently, 
the interest in the field of the low-dimensional quantum systems
is concentrated on systems with a spin gap, 
especially
in connection with the high-$T_{\rm c}$ cuprates upon doping.\cite{dagotto-r96}
The ground state of such a system realizes the spin disordered singlet state,
favored by a quantum fluctuation and/or frustration.
The excitation energy gap opens above the ground state.
As typical models, we can consider
the spin ladder model,\cite{ladders,alters} 
the bond-alternation model,\cite{alters} 
and the Majumdar-Ghosh model.\cite{majumdar-g69,shastry-s81}

Syntheses of various new compounds \cite{ramirez94}
also accelerate investigations on this field both theoretically 
and experimentally.
For example, magnetic susceptibility measurement on KCuCl$_3$ 
\cite{tanaka-tso96} indicates a spin gap behavior, 
and the experimental data are considered to be explained theoretically
by the double spin-chain model with frustration. \cite{nakamura-ot96}
Estimation of the magnitude of the gap through the susceptibility data
is already established on the assumption of the dimer gap.\cite{troyer-tw94}
However, it is not sufficient to the thorough understandings of the system.
We must investigate the origin of the gap and the 
physical picture of the excited states.
This is a main purpose of this paper.

The $\Delta$ chain, the subject model in the present paper, 
is a participant of double spin-chain systems, but
has rather special geometry of the interaction bonds.
The triangles are aligned in one direction.
If we connect the spins located at the top of the triangles with the 
interaction bonds, it becomes the railroad-trestle model.
Therefore, it may seem that the interesting 
properties are peculiar to this special system.

The early investigations on the $\Delta$ chain were directed in search 
for a system with the singlet dimer ground state.
\cite{long-f90,long-s90,monti-s91,monti-s92}
Interpretation of the excited states by a kink and an antikink was 
done at this stage.\cite{long-f90,long-s90}
The existence of the finite energy gap above the ground state was
rigorously proven.\cite{monti-s92}
Analogy of the model to the kagom\'e antiferromagnet
has been pointed out:\cite{kubok93} 
there are macroscopic local continuous degeneracy in the ground state 
in the classical limit,
low-lying excitation spectrum is consequently dispersionless, and 
there exists an additional peak of the specific heat at low temperatures.
The origin of the dispersionless mode and the double peak of the specific heat
were recently clarified.\cite{nakamura-k96,sen-swc96}
Sen {\it et al}\cite{sen-swc96} also pointed out 
that the possible relevance of the model 
to the newly synthesized compound, YCuO$_{2.5}$.

In this paper, we find that the $\Delta$ chain
possesses common features with other spin gap systems besides its
unique properties stated above.
Concretely, we mean the ``common feature'' by the way how the system
reduces frustration by the bond dimerization.
This is an essence of this paper.

The ground state is generally unstable against a small perturbation 
that relaxes strong frustration.
It must have a strong influence on the low-lying excited states as well, 
and consequently to the low-temperature behavior of 
various physical quantities.
We should take this effect into account when we analyze the 
experimental data.
We consider the bond dimerization as a perturbation, 
since its energy stabilization is the strongest one, and
it can be realized by a lattice distortion as in the spin-Peierls systems.
\cite{okamoto-nt86}

From a theoretical point of view, the frustrated double spin-chain systems
with the bond dimerization are very attractive.
They have the dimer state and the $S=1$ Haldane state in both extremes of
the strength of the dimerization parameter.
In the midst is the fully-frustrated point.
As was shown in the bond-alternation model,\cite{alters}
there always exist the dimer order \cite{hida92} and the string order of 
den Nijs and Rommelse.\cite{dennijs-r89}
The dimer state continuously changes to the $S=1$ Haldane state
with the increase of strength of the ferromagnetic interaction bonds.
Therefore, there is always a finite energy gap.
Then questions arise.
How do we identify these two phases?
How the differences manifest in the observable physical quantities?
In this paper, we show by
adopting the $\Delta$ chain for an example
that the whole phase space can be divided into three regions with
respect to the nature of the first excited state relevant to the 
energy gap,
and that the differences affect the low-temperature dependences of 
physical observables such as the specific heat and the susceptibility.

Section \ref{sec:model} describes the model and summarizes 
the general remarks concerning the symmetric 
$\Delta$ chain without the dimerization,
which will be the starting point of the discussion in Sec. \ref{sec:near1}.
%
%We also show a simple numerical result of the gap behavior with 
%respect to the dimerization.
%
In the first part of Sec. \ref{sec:near1},
we do the same variational analysis as was done in the 
symmetric $\Delta$ chain.\cite{nakamura-k96,sen-swc96}
This is valid only for positively (antiferromagnetically) small dimerization.
Then,
continuous limit of the effective Hamiltonian is derived and its 
exact solution is obtained.
The remaining part of Sec. \ref{sec:near1} is devoted to the case when 
the dimerization is negatively (ferromagnetically) small.
We use the non-local unitary transformation 
\cite{kennedy-t92,takada-k91,takada92,hida-t92}
to represent the ground state and the excited state.
This transformation is 
equivalent to the Kennedy-Tasaki transformation of the $S=1$ system,
and is its adaptation to the double spin-chain systems.
It is known to be powerful when the ground state is either
in the Haldane state or in the state with strong dimer correlation.
Almost equivalent results to the positive dimerization case are obtained
near the symmetric point.
We show how the system converges to the Haldane state
and the dimer state in Sec. \ref{sec:away1}.
Section \ref{sec:obs} shows the differences of the observable quantities
between two phases.
We also propose a quantity to judge the phase in experiments.
Section \ref{sec:sum} is devoted to summary and discussion.

%%%%%%%%%%%%%%%%%%%%%%%%%%%%%%%%%%%%%%%%%%%%%%%%%%%%%%%%%%%%%%%%%%%%%%%%%%%%
\section{The model and general remarks}
\label {sec:model}
\begin{figure}[h]
 \epsfxsize = 7.5cm
 \epsffile{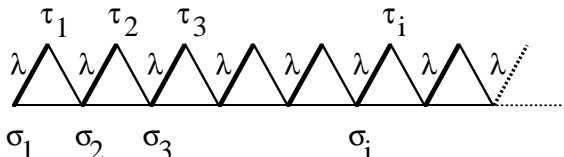}
 \caption {Shape of the dimerized $\Delta$-chain.
           Bold lines indicate the $\lambda$ bonds.
  \label{fig:lattice}
          }
\end  {figure}

%%%%%%%%%%%%%%%%%%%%%%%%%%%%%%%%%%%%%%%%%%%%%%%%%%%%%%%%%%%%%%%%%%%%%%%%%%%%
We consider a system described by the following Hamiltonian.
\begin{equation}
 {\cal H}=\sum_{n=1}^N 
    \lambda\mbox{\boldmath $\sigma$}_n \cdot\mbox{\boldmath $\tau  $}_{n}
         + \mbox{\boldmath $\tau  $}_n \cdot\mbox{\boldmath $\sigma$}_{n+1}
         + \mbox{\boldmath $\sigma$}_n \cdot\mbox{\boldmath $\sigma$}_{n+1}
\end  {equation}
Here, $N$ is the number of the triangles in the system, $\lambda$ is 
a parameter denoting the dimerization, 
and $|\mbox{\boldmath $\sigma$}|=|\mbox{\boldmath $\tau$}|=1/2$.
Figure \ref{fig:lattice} shows the depicted lattice.
At the point of $\lambda=1$, the system is the symmetric
$\Delta$ chain and is fully frustrated.
In the limit of $\lambda=+\infty$, the system reduces to the independent
dimer model.
In the other limit of $\lambda=-\infty$, the system becomes
equivalent to the $S=1$ spin chain.
Therefore, the present model has the independent dimer ground state 
and the Haldane ground state in its extremes.

The understandings of the symmetric $\Delta$ chain
are already established.\cite{nakamura-k96,sen-swc96}
Here, we briefly summarize the results.
The ground state is the perfect singlet dimer state with two-folded degeneracy
in the case of the periodic boundary conditions.  
The low-lying excited states approximately consist of 
$(N-1)$ singlet dimers and two free spins.
The excitations are governed by sets of two free spins named 
a ``kink'' and an ``antikink''.
A kink stays localized and works as a delimiter to moving antikinks.
Dispersionless aspects of the excitations originate in a localized kink.
An antikink is considered as a free particle moving between kinks 
with the effective mass.
An antikink is supposed to be one free spin only 
within the first approximation,
since it is not an eigenstate of the local Hamiltonian.
It spreads out to an extent in reality.
A detailed structure of an antikink is not revealed yet, but 
it merely renormalizes the effective mass.
Within the first approximation, the mass $m=1$.
The second approximation that an antikink is distributed among 5 spin sites
gives the mass $m=1.158$,\cite{sen-swc96} and
the numerical diagonalization data after the 
$N\to\infty$ extrapolation shows the mass $m=1.21$.\cite{nakamura-k96}
The excitation gap is well expressed by a summation of a kinetic energy of an 
antikink and the creation energy of a pair of a kink and an antikink, 
$\epsilon_0 = 0.215$, as
\begin{equation}
  \Delta E=\epsilon_0 + \frac{1}{m}\left (1-\cos \frac{\pi k}{N}\right ),
\end  {equation}
where $k$ is the wave number of an antikink alone.
The low-temperature peak of the specific heat can be
reproduced by the Schottky specific heat with the gap expressed above.
\cite{nakamura-k96}
Susceptibility was also calculated.\cite{sen-swc96}

%%%%%%%%%%%%%%%%%%%%%%%%%%%%%%%%%%%%%%%%%%%%%%%%%%%%%%%%%%%%%%%%%%%%
\section{In the vicinity of $\lambda=1$}
\label {sec:near1}

We numerically diagonalized the above Hamiltonian up to the systems with 
28 spins $(N=14)$ under the periodic boundary conditions. 
\begin{figure}[h]
 \epsfxsize = 8cm
 \epsffile{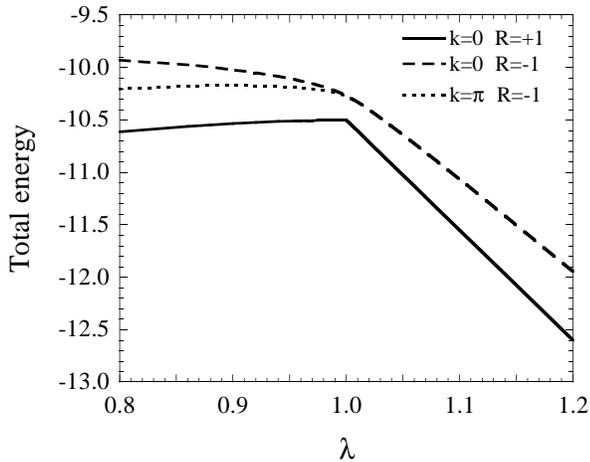}
 \caption{The $\lambda$ dependence of the energy of the
          lowest state in the subspace denoted by $k$ (wave number), 
          and $R$ (spin reversal symmetry) in the system with $N=14$ 
          (28 spins).
 \label{fig:diagE} 
         }
\end  {figure}
The $\lambda$ dependence of the energy is shown in Fig. \ref{fig:diagE}.
The ground state energy is highest 
at the fully-frustrated point ($\lambda=1$),
which leads to the instability due to the lattice dimerization.

For $\lambda > 1$, the ground state is always the pure dimer state
which consists of the singlet dimers on the every $\lambda$-bond.
Thus the total ground state energy is exactly $-0.75\lambda N$.
We refer to this state as the left-dimer state hereafter in this paper.
In this region, the spin correlation length vanishes.
The excited states also have no $k$-dependence.
At $\lambda = 1$, the ground states are two-fold degenerate under the
periodic boundary conditions.
They are the left-dimer state and the right-dimer state.
The first-order transition occurs at this point.
For $\lambda < 1$, the correlation length gradually increases with 
decreasing $\lambda$.
The right-dimer state smoothly reaches the $S=1$ Haldane state
in the limit of $\lambda = -\infty$ without any phase transition.
Hereafter, we simply call the region $\lambda > 1$ as the dimer phase, 
and the region $\lambda < 1$ as the Haldane phase.
The degeneracy of the excited states in the dimer phase is
lifted in the Haldane and a state with $k=\pi$ becomes the first excitation.

%%%%%%%%%%%%%%%%%%%%%%%%%%%%%%%%%%%%%%%%%%%%%%%%%%%%%%%%%%%%%%%%%%%%
\subsection{$\lambda > 1$: the dimer phase}
The results obtained in this subsection has been reported briefly.
\cite{nakamura-t96}
We discuss the method and the results in detail in this paper.

The excited states in this phase may have similar properties to those
at the symmetric point, because the ground state is the same.
Therefore,
we do the same variational analysis 
as was successful at $\lambda =1$.\cite{nakamura-k96,sen-swc96}
The boundary conditions are set open, though it does not influence the 
final results in the thermodynamic limit.
We define a variational basis $\psi_i$ 
so that an antikink is located at the $i$th triangle.
\begin{eqnarray}
 \psi_i \equiv  \psi_{\rm kink}&\otimes& [4,5] \cdots [2i-2,2i-1] 
                \uparrow_{2i} 
                \nonumber \\
       &      &         [2i+1, 2i+2] \cdots [2N-1, 2N],
\end  {eqnarray}
where $[i, j]$ denotes a singlet dimer state connecting the $i$th and 
the $j$th site,
namely $[i, j]=(\uparrow_i\downarrow_j-\downarrow_i\uparrow_j)/\sqrt{2}$ for
$\uparrow_i (\downarrow_i)$ denoting an 
up (down) spin located at the $i$th site.
Here, we numbered the site so that $\mbox{\boldmath $\sigma$}_n$ is the
$(2n-1)$th, and $\mbox{\boldmath $\tau$}_n$ is the $(2n)$th
as is shown in Fig. \ref{fig:lattice}.
The $\uparrow_{2i}$ above is an antikink.
The wave function of a kink located at the leftmost edge is known as,
\begin{equation}
 \psi_{\rm kink}=[\uparrow_1(\uparrow_2\downarrow_3-\downarrow_2\uparrow_3)
                 +\uparrow_2(\uparrow_1\downarrow_3-\downarrow_1\uparrow_3)]
                 /\sqrt{6}.
\end  {equation}
This state is an eigenstate of the edge triangle Hamiltonian, and therefore
does not move.
Its energy eigenvalue is $\lambda /4-1$.
Thus a kink contributes to the excitation energy by $(\lambda -1)$.
The singlet dimers, $\{[2n, 2n+1]\}$, existing between 
a kink and an antikink, are not eigenstates of local 
triangle Hamiltonians.
They also contribute to the excitation energy.

A variational basis is not orthogonal to each other and satisfy the following 
relations.
\begin{eqnarray}
 \langle \psi_i  |\psi_j\rangle&=&\left ( \frac{1}{2}\right )^{|i-j|},
                               \label{eq:overlap}\\
 \langle \psi_i|{\cal H}|\psi_j\rangle&=&\left[E_{\rm g}
                                 +(\lambda-1) 
                                 +\frac{3}{4}(\lambda-1)\min (i,j) \right]
                                  \langle \psi_i|\psi_j\rangle \nonumber \\
                                 &+&\frac{3}{4}\delta_{ij}.
                               \label{eq:hamil}
\end  {eqnarray}
Here, $\delta_{ij}$ is the Kronecker delta, and $\min (i,j)=i$ if $i\leq j$.
Our task is to find a function $\Psi_{\rm var}\equiv \sum_{i} C_i\psi_i$
that minimizes the energy expectation,
 \begin{equation}
   {\rm Var}E\equiv 
   \frac{\langle \Psi_{\rm var} |{\cal H}|
                 \Psi_{\rm var} \rangle}
        {\langle \Psi_{\rm var} |
                 \Psi_{\rm var} \rangle}
   = 
   \frac{\sum_{i,j}C_i C_j \langle \psi_i|{\cal H}|\psi_j\rangle}
        {\sum_{i,j}C_i C_j \langle \psi_i|         \psi_j\rangle}.
 \label {eq:vare}
 \end  {equation}

\begin{figure}[h]
 \epsfxsize = 8cm
 \epsffile{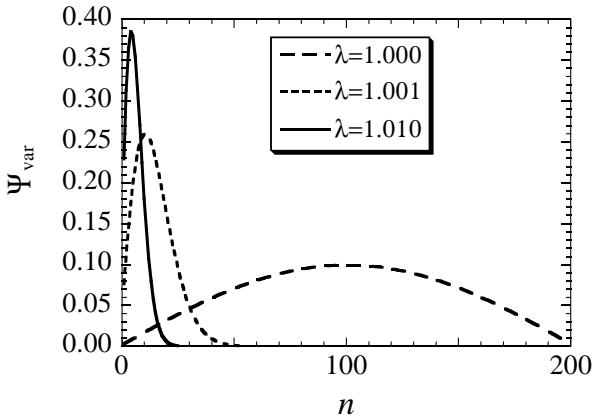}
 \caption {Variational wave function of an antikink in the lowest excited state 
           for $\lambda=1.00, 1.001$, and $1.01$.
           Size of the system $N=200$.
           $n$ stands for the site number of the triangles.
 \label{fig:wavef}
          }
\end  {figure}

If we diagonalize the denominator and rewrite the numerator with 
its eigenfunction,
this variational problem is transformed into a simple eigenvalue problem.
Then, the solution of Eq. (\ref{eq:vare}) can be obtained by the numerical
diagonalization for a finite system size.
We show the result of $N=200$ in Fig. \ref{fig:wavef} with
$\lambda = 1.00, 1.001$ and $1.01$.
In the case of the symmetric $\Delta$ chain ($\lambda =1.00$), 
the wave function is the 
sine function indicating a free motion of an antikink.
An antikink is drastically attracted to a kink as $\lambda$ increases.
The wave function exhibits an antikink localized.
We clarify the wave function analytically by using the continuous limit.

We rewrite the above equation by the new basis 
$|\phi_k\rangle=\sum_{n}\exp[ikn]|\psi_n\rangle$,
since the denominator of Eq. (\ref{eq:vare}) 
is diagonalized by Fourier transformation 
in the large $N$ limit.
Note that $k$ is the wave number of an antikink alone, and
does not correspond to the total wave number.
Then the basis relations become
\begin{eqnarray}
 \langle \phi_k  |\phi_l\rangle&=&
      \frac{3}{5-4\cos k}\delta_{k,l}       %\nonumber \\
      \label{eq:koverlap}
 \\
 \langle \phi_k|{\cal H}|\phi_l\rangle&=&
     \left[E_{\rm g} +(\lambda-1) \right]\langle \phi_k|\phi_l\rangle
     +\frac{3}{4}\delta_{k,l} \nonumber \\
   &+&\frac{3}{4}(\lambda-1) A_{k,l},
      \label{eq:kmatrix}
\end  {eqnarray}
with 
\begin{equation}
 A_{k,l}=\frac{1}{N}\sum_{n,m}\exp[i(kn-lm)]\min (m,n)
          \langle\psi_n|\psi_m\rangle.
\end  {equation}
The diagonal elements of $A_{k,l}$ dominate the off-diagonal elements.
Then, $A_{k,k}$ is written as
 \begin{eqnarray}
   A_{k,k}&=&
            \frac{1}{N}
            \frac{6-13\cos k +8\cos 2k}
                 {\frac{245}{4} -87\cos k + 30\cos 2k + 4\cos 3k}\nonumber \\
         &+&
             \frac{4-5\cos k}
                {\frac{33}{4}-10\cos k +2\cos 2k}\nonumber \\
          &+&\frac{N+1}{2}\frac{3}{5-4\cos k}.
          \label{eq:Akk}
 \end  {eqnarray}
We find in this equation that
$A_{k,k}$ diverges in the limit $N\to\infty$.
In order to avoid this divergence,
we introduce a cut-off factor $i\delta$ to the momentum,
namely $k\to k+i\delta/2$.
This should have no effect on physical results by taking 
the limit $\delta\to 0$ after $N\to \infty$.
We rewrite the matrix element and pick only up the 
leading term of the off-diagonal part and the terms 
up to the $k^2$ in the diagonal part.
Then the continuous limit, $N\to\infty$ and $k\to 0$, 
of the Hamiltonian $\tilde {\cal H}_{k,l}$
is given by,
\begin{eqnarray}
 \tilde {\cal H}_{k,l}&\doteq& 
        \left(E_{\rm g}+(\lambda-1)+ \frac{1}{4}+\frac{k^2}{2}\right)
        \delta_{k,l}
        \nonumber \\
     &+&\frac{1}{N}\frac{3}{4}(\lambda-1)
        \frac{1}{(k-l+i\delta)^2}.
\end  {eqnarray}
Apart from the constant terms, this Hamiltonian is equivalent to the 
following in the real space representation.
\begin{equation}
 {\cal H}_{C}= -\frac{1}{2m}\frac{d^2}{dx^2}
             +\frac{3}{4}(\lambda-1)x \exp[-\delta x].
\end  {equation}
Here, $x$ is the distance between a kink and an antikink;
$m$ is the effective mass of an antikink and is set $m=1$ in the 
first approximation.
We take the limit $\delta\to 0$ at this stage.
The first term is the kinetic energy of an antikink, and 
the second term is the triangular potential attracted by a localized kink.
We rescale $x$ by $X=\theta x$ with 
$\theta = (3m(\lambda-1)/2)^{-1/3}$.
Then the eigenvalue equation ${\cal H}_{\rm C} \Psi = E_{\rm C} \Psi$ becomes
\begin{equation}
\left (-\frac{d^2}{dX^2} + X\right )\Psi = E'\Psi,
\end  {equation}
where $E'=E_C \times 2m\theta ^2$.
Its solution is known as the Airy function,
$\Psi = A_i(X-E')$ with the first eigenvalue 
$E'\simeq (3\pi/2 \times 0.7587)^{2/3} \simeq 2.338$.
\cite{stern72}
Accordingly, we can obtain the energy eigenvalue $E_C$, 
the average distance between a kink and an antikink $\langle x\rangle$,
and the localization length of the wave function $\xi$.
\begin{eqnarray}
  E_C&=&\frac{E'}{2m\theta ^2}=1.532m^{-1/3}(\lambda -1)^{ \frac{2}{3}}.
     \label {eq:ec1} \\
 \langle x \rangle &=&
         \frac{2}{3}E' \theta =1.362m^{-1/3}(\lambda -1)^{-\frac{1}{3}}.
     \label {eq:x1} \\
  \xi&\sim& 5 \times \theta   =4.368m^{-1/3}(\lambda -1)^{-\frac{1}{3}}.
     \label {eq:xi1}
\end  {eqnarray}
The localization length,$\xi$, 
is obtained from a rough estimation of the localization length of the
Airy function $\sim 5$.
The estimation of $\xi$ is quite consistent with the wave function 
shown in Fig. \ref{fig:wavef}:
$\xi\sim 40$ for $\lambda=1.001$, and $\xi\sim 20$ for $\lambda=1.01$.
It should be noticed that
$\frac{3}{4}A_{k,k}/\langle\phi_k|\phi_k\rangle$ of Eq. (\ref{eq:koverlap})
and (\ref{eq:kmatrix}) becomes equivalent to $E_{\rm C}$,
if we replace $N$ by the $\xi$ above.

\begin{figure}[h]
 \epsfxsize = 8cm
 \epsffile{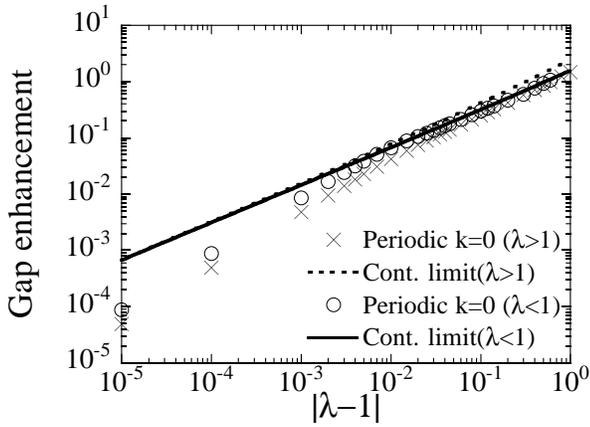}
 \caption {Log-log plot of the gap enhancement for 
           the exact result in the continuous limit with the mass $m=1.21$,
           and the numerical results  of the periodic system with $N=14$.
 \label{fig:logloggap}
          }
\end  {figure}

Now, 
the total gap behavior can be obtained from the kink contribution, 
$(\lambda -1)$, and the antikink contribution, $E_C$. 
The gap enhancement defined by 
$\Delta _{\rm Gap}=E_{\rm gap}(\lambda)-E_{\rm gap}(1)$ is
\begin{equation}
 \Delta _{\rm Gap}=(\lambda -1) + 1.5319\times m^{-1/3} (\lambda -1)^{2/3}.
\end  {equation}
The gap increases in a power law with its exponent $2/3$.
Figure \ref{fig:logloggap} shows the behavior of the gap enhancement,
$\Delta _{\rm Gap}$,
compared with the numerical results of the periodic system of $N=14$.
We used a value of the mass, $m=1.21$, estimated at $\lambda=1$ 
\cite{nakamura-k96} for better comparison.
For $\lambda -1 > 0.01$, the numerical data agree with the analytical 
estimation.
In this region, the localization length is within the finite 
system size, $\xi < 14$.

The relevant excitation in this region is governed by a competition between 
the kinetic energy and the potential energy of an antikink.
When the dimerization is small, an antikink gains energy by the 
kinetic motion.
As the dimerization becomes large, an antikink is bound by a kink, 
and finally they collapse to a local triplet.
The exponent $2/3$ is a general outcome of this competition.

\subsection{$\lambda < 1$: the Haldane phase}

In this region, the ground state cannot be known trivially,
although we can expect the right-dimer state continuously changes to the 
$S=1$ Haldane state in the limit of 
$\lambda \to -\infty$.\cite{hida92,takada92,hida-t92}
We make use of the non-local unitary (NLU) transformation for the 
double spin-chain systems,\cite{kennedy-t92,takada-k91,takada92,hida-t92} 
the second-order perturbation, 
and the numerical diagonalization to clarify the ground state first.
Then we proceed to investigate the excited state.

We transform the Hamiltonian ${\cal H}$
with $U$ defined in Appendix \ref{app:nlu}.
\begin{eqnarray}
  U^{-1}{\cal H}U =
    \sum_{n=1}^N &&
  \lambda \mbox{\boldmath $\sigma$}_n \cdot \mbox{\boldmath $\tau  $}_n
    \nonumber \\
   &-& (\sigma_n^x+\tau_n^x)\tau  _{n+1}^x
    -  (\sigma_n^z+\tau_n^z)\sigma_{n+1}^z   \nonumber \\
   &-&4(\sigma_n^z \tau_n^x + \sigma_n^x \tau_n^z)\sigma_{n+1}^z\tau_{n+1}^x.
   \label{eq:transhamil}
\end  {eqnarray}
According to this transformation,
string order parameters are transformed into local parameters such as
$\langle \sigma_n^x \rangle$, $\langle \sigma_n^z\rangle$,
$\langle \sigma_n^z \tau_n^x \rangle$,
as will be found in Eqs. (\ref{eq:odim}) and (\ref{eq:ostr}).
It makes possible that we employ a single-site approximation like the 
molecular-field one.
Thus,
we consider the following variational trial function for the ground state.
\begin{eqnarray}
 |\Psi_0\rangle &=&\prod_{n=1}^N |n(b)\rangle =
          \prod_{n=1}^N(b|T_n\rangle + \sqrt{1-b^2}|S_n\rangle) 
    \label{eq:varbase}               \\
 |S_n\rangle &=&(|\uparrow, \downarrow\rangle 
                -|\downarrow, \uparrow\rangle)/\sqrt{2} \\
 |T_n\rangle &=&\alpha |\uparrow,   \uparrow  \rangle
               +\beta (|\uparrow,   \downarrow\rangle 
                      +|\downarrow, \uparrow  \rangle)/\sqrt{2}
               +\gamma |\downarrow, \downarrow\rangle
\end  {eqnarray}
$|\uparrow, \uparrow\rangle$'s are the state of $|\sigma_n^z, \tau_n^z\rangle$.
$b, \alpha, \beta, \gamma$ are the variational parameter and satisfy 
the normalization condition, $\alpha ^2 + \beta ^2 + \gamma ^2 =1$.
These parameters are supposed to be invariant of $n$.
The analysis is variational in this sense.
A state with $b=0$ is the left-dimer state on the $\lambda$ bonds,
a state with $b=\sqrt{3}/2$ is the right-dimer state,
and
a state with $b=1$ corresponds to the pure VBS state in this representation.
The energy expectation value is calculated as
\begin{eqnarray}
  \langle \Psi_0|{\cal H}| \Psi_0 \rangle
   &=& \Biggl [ \lambda \left( b^2-\frac{3}{4} \right ) \nonumber \\
   &-& b^4\left ( \frac{(\alpha^2-\gamma^2)^2}{2}
                 +2\beta^2(\alpha^2+\gamma^2)\right ) 
\nonumber \\
    &-&3b^3\sqrt{1-b^2}\beta (\alpha^2-\gamma^2)\Biggr ] N.
\end  {eqnarray}
We can easily find this minimum value and the variational parameter set 
by using the Lagrange multiplier.
The energy value $\epsilon_0$ is
\begin{equation}
 \lambda \left ( b^2-\frac{3}{4} \right ) -\frac{2}{3}b^4 
  -\frac{2b^3\sqrt{3(1-b^2)}}{3} \equiv \epsilon_0,
\end  {equation}
with four possible choices of the parameters $(\alpha, \beta, \gamma)$ as
\begin{eqnarray}
(\alpha, \beta, \gamma)&=&(\pm \sqrt{2/3}, \sqrt{1/3}, 0),\nonumber \\
                        &&(0, -\sqrt{1/3},\pm \sqrt{2/3}),
 \label{eq:abg}
\end  {eqnarray}
and $b$ determined through
\begin{equation}
  \lambda = \frac{4}{3}b^2-\frac{b(4b^2-3)}{\sqrt{3(1-b^2)}},
  \label{eq:lamdet}
\end  {equation}
or
\begin{equation}
  b=0.
\end  {equation}
The four-fold degeneracy in the choice of $\alpha, \beta, \gamma$
corresponds to the degeneracy of the edge states.
\cite{takada-k91,takada92,fath-s93}
A state with $b=0$ is the singlet dimer ground state for $\lambda > 1$.
The other one, Eq. (\ref{eq:lamdet}), 
corresponds to the ground state for $\lambda < 1$.
If we solve Eq. (\ref{eq:lamdet}) up to the second order of $(1-\lambda)$,
\begin{equation}
  b=\frac{\sqrt{3}}{2}\left[1+\frac{1-\lambda}{4}-\frac{5}{16}(1-\lambda)^2
                      \right ].
\end  {equation}
Then the energy expectation per triangle is
\begin{equation}
  E/N=-\frac{3}{4}-\frac{3}{16} (1-\lambda)^2.
  \label{eq:vareg}
\end  {equation}
This agrees with the result due to
the second-order perturbation of 
$(\lambda-1) \mbox{\boldmath $\sigma$}_n \cdot \mbox{\boldmath $\tau$}_n$.
Details of the calculation are given in Appendix \ref{app:2ndp}.
The diagonalization result of a finite system with 14 triangles 
shown in Fig. \ref{fig:diagE} is fitted by
the least-square method to
\begin{equation}
  E/N=-\frac{3}{4}-0.1886 (1-\lambda)^2-0.003 (1-\lambda).
\end  {equation}
Consistency between the analytic estimation (note $3/16=0.1875$)
and the numerical result is excellent.

We can also estimate the string order parameter of den Nijs and Rommelse,
$O_{\rm str}$, \cite{dennijs-r89}
and the dimer order parameter,$O_{\rm dim}$. \cite{hida92,takada92}
The definitions and the expectation values are
\begin{eqnarray}
 O_{\rm dim}&=&\lim_{|m-n|\to\infty}-4\langle U^{-1}\tau_m^z
             \exp  \left[i\pi \sum_{k=m+1}^{n-1} S_k^z \right ]
             \sigma_n^z U \rangle                  \nonumber \\
         &=&\lim_{|m-n|\to\infty} 4\langle \sigma_m^z \sigma_n^z \rangle
          = 4\langle \sigma_m^z \rangle \langle \sigma_n^z \rangle
          = 4\langle \sigma_m^z \rangle ^2  \nonumber \\
            &=&\frac{4}{9}b^4+\frac{4}{3}b^2(1-b^2)
              +\frac{8}{9}b^3\sqrt{3(1-b^2)},   \label{eq:odim} \\
 O_{\rm str}&=&\lim_{|m-n|\to\infty} -\langle U^{-1} S_m^z
             \exp      [i\pi \sum_{k=m+1}^{n-1} S_k^z       ]
             S_n^z U\rangle                        \nonumber \\
         &=&\lim_{|m-n|\to\infty} \langle S_m^z S_n^z \rangle
          = \langle S_m^z \rangle \langle S_n^z \rangle
          = \langle S_m^z \rangle ^2 \nonumber \\
            &=&\frac{4}{9}b^4.
   \label{eq:ostr}
\end  {eqnarray}
At $\lambda =1$, $O_{\rm dim}=1$ as $b=\sqrt{3}/2$.
In the limit of $\lambda\to -\infty$, it converges 
to the VBS value $O_{\rm dim}=4/9$ as $b\to 1$.
On the other hand, the string order parameter, $O_{\rm str}$ takes a value
of $O_{\rm str}=1/4$ at $\lambda=1$ and converges to the same value of 
$O_{\rm dim}=O_{\rm str}=4/9$.
It should be noted that the convergence of $b\to 1$ is very fast that
it takes a value $b=0.94$ even at $\lambda = 0.25$.
Therefore, the Haldane state can be realized even if all the 
interaction bonds are antiferromagnetic.

\begin{figure}[h]
 \epsfxsize = 8cm
 \epsffile{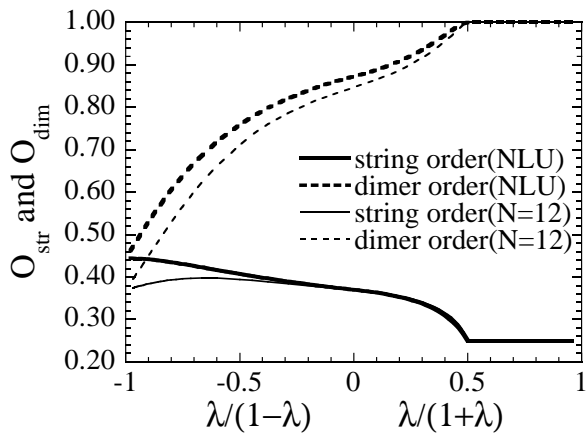}
 \caption {The string order parameter and the dimer order parameter are plotted
           against $\lambda/(1-\lambda)$ for $\lambda < 0$, and 
           against $\lambda/(1+\lambda)$ for $\lambda >0$.
           Bold lines are the analytic estimate obtained by the non-local 
           unitary transformation.
           Thin lines are numerical diagonalization results of the lattice 
           with 12 triangles.
 \label{fig:str}
          }
\end  {figure}

Figure \ref{fig:str} shows the dimer and the string order parameter of the 
ground state in the whole phase space.
Numerical data are of the periodic system with $N=12$.
Quantitative agreement is excellent near $\lambda=1$.
The string order increases as $\lambda$ decreases from 1.
Therefore, we call the region $\lambda < 1$ the Haldane phase.
It should be noted that
the numerical 
data converges to the value of $S=1$ ($\simeq 0.37$)\cite{white-h93}
in the limit of $\lambda \to -\infty$, not to the value of the pure VBS state.
The contradiction becomes distinct at $\lambda \sim -1$,
where the string order parameter takes the maximum value.
This is because we used the variation of the single-site approximation
as in Eq. (\ref{eq:varbase}), besides
the correlation length is rather large in the Haldane state.

A low-energy excitation is intrinsically of domain-wall type in one dimension.
Two perfect singlet dimer states, $|n(\sqrt{3}/2)\rangle$ and
$|n(0)\rangle$, are degenerate at $\lambda=1$. 
For $\lambda < 1$, the latter state becomes the excited state and 
the former state becomes the ground state by changing the value of $b$.
Therefore,
we consider the following wave function as
a trial for the excited state.
\begin{equation}
 |\Psi_1\rangle = \sum_i C_i \psi_i 
                = \sum_i C_i 
                 \left (
                \prod_{n=  1}^i| n(b)\rangle 
                \prod_{n=i+1}^N| n(0)\rangle
                \right ).
\end  {equation}
Because the state $| n(0)\rangle $ becomes a higher excited state for
$\lambda < 1$, the validity of this trial wave function is 
restricted to the very vicinity of $\lambda =1$.
The domain wall located at the $i$th triangle is essentially an antikink
before the NLU transformation is done.
Thus, the analysis is for a kink-antikink excitation.
The basis relations of $\psi_i$ are 
\begin{eqnarray}
  \langle\psi_i|\psi_j\rangle &=& (\sqrt{1-b^2})^{|i-j|}  \\
  \langle\psi_i|{\cal H_o}|\psi_j\rangle &=&
     (E_{\rm g}+E_1 (N-\min(i,j))+E_2\delta_{i,j}) \nonumber \\
                           &&  \langle\psi_i|\psi_j\rangle,
\end  {eqnarray}
with $ E_{\rm g}=\epsilon_0 N$,
     $ E_1      =-0.75 \lambda - \epsilon_0$, and
     $ E_2      = \lambda (b^2-0.75)-\epsilon_0$.
These relations are equivalent to the ones in the previous subsection,
Eq. (\ref{eq:overlap}) and (\ref{eq:hamil}).
Therefore, the problems are solved in the same way and we obtain 
\begin{eqnarray}
  E_C               &=   & 1.856m^{-1/3}E_1^{ \frac{2}{3}},\label{eq:ec2}\\
  \langle x \rangle &=   & 1.237m^{-1/3}E_1^{-\frac{1}{3}},\label{eq:x2}\\
  \xi               &\sim& 3.969m^{-1/3}E_1^{-\frac{1}{3}},\label{eq:xi2}
\end  {eqnarray}
for the excitation of $k=0$.
However, we remark that the lowest excitation is the state 
with $k=\pi$ in this region.
This state is considered as that both a kink and an antikink 
are mobile with the total momentum $\pi$.
The analysis on the state of $k=\pi$ will be done in the next subsection.
Three estimates, Eqs. (\ref{eq:ec2}), (\ref{eq:x2}), and (\ref{eq:xi2})
above, however,
have quite good consistency with the second excited state with $k=0$.

\section{Away from $\lambda=1$}
\label {sec:away1}
Away from the fully-frustrated point, $\lambda =1$, the excitation is no longer
of a kink-antikink type.
They already collapse to a local triplet state.
Therefore, we must consider another type of a domain wall for 
a trial wave function.

In the region of $\lambda < 1$, 
the ground state has four-fold degeneracy associated with 
the edge states.
Within the scheme of our variational analysis, it appears in the 
four possible choices of the variational parameter 
$(\alpha, \beta, \gamma)$ of Eq. (\ref{eq:abg}).
The most natural candidate for the domain wall is 
the one between any two of the four-fold degenerate ground states.
In fact, F\'ath and J. S\'olyom \cite{fath-s93} showed that the 
lowest excitation in the AKLT model \cite{aklt}
is of this type.
The trial wave function we consider is then 
\begin{equation}
 |\Psi_1\rangle = \sum_i C_i \psi_i 
                = \sum_i C_i 
                 \left (
                \prod_{n=  1}^i| n(b)\rangle 
                \prod_{n=i+1}^N| n'(b)\rangle
                \right ),
\label{eq:varex}
\end  {equation}
where $|n(b)\rangle$ and $|n'(b)\rangle$ only differ the choice of the 
set $(\alpha, \beta, \gamma)$.  
We use the set
$(\alpha, \beta, \gamma)=(\sqrt{2/3},\sqrt{1/3},0)$
for $|n(b)\rangle$,
and $(\alpha, \beta, \gamma)=(-\sqrt{2/3},\sqrt{1/3},0)$
for $|n'(b)\rangle$. 
$b$ is common and determined by Eq. (\ref{eq:lamdet}).
The basis relations are calculated in the same way as
\begin{eqnarray}
  \langle \psi_i|\psi_j\rangle &=& \left (1-\frac{4}{3}b^2\right )^{|i-j|} 
                               \equiv (-a)^{|i-j|},\\
  \langle \psi_i|{\cal H}|\psi_j\rangle &=&
  [E_{\rm g} + (|i-j|-1) E_1 + \delta_{i,j} (E_1+E_2)] \nonumber \\
  &&\langle \psi_i|\psi_j\rangle,
\end  {eqnarray}
with
\begin{eqnarray}
  E_1&=&-\frac{2}{3}b^3\left (b+\sqrt{3(1-b^2)} 
                       + \frac{3a}{(b+\sqrt{3(1-b^2)})^3}\right),
  \label{eq:e1}\\
  E_2&=&\frac{8}{9}b^3(b+\sqrt{3(1-b^2)}).
  \label{eq:e2}
\end  {eqnarray}
In the thermodynamic limit, the above matrices are diagonalized by the
Fourier transformation.
An expectation value of the energy gap above the ground state is 
estimated as
\begin{eqnarray}
  E_{\rm ex}(k)&=&
         \frac{\langle \phi_k|{\cal H}|\phi_k\rangle}
              {\langle \phi_k|         \phi_k\rangle}-E_{\rm g}
          \nonumber \\
          &=&
         -E_1 \left( 
         1+\frac{2a}{1-a^2}
           \frac{(1+a^2)\cos k + 2a}{1+2a\cos k + a^2}  \right.\nonumber \\
         && \left. -\frac{1+2a\cos k + a^2}{1-a^2}  \right)
               +E_2
           \frac{1+2a\cos k + a^2}{1-a^2}.
  \label{eq:gaplocal1}
\end  {eqnarray}
$E_{\rm ex}(k)$ always takes minimum at $k=\pi$.
The energy gap $E_{\rm ex}(\pi)$ converges to $1/9$ in the VBS limit,
$\lambda \to -\infty$.

We can also discuss excited states of $\lambda > 1$ with this 
variation scheme, since
the domain wall of Eq. (\ref{eq:varex}) can express the local triplet 
domain wall, if we change the notation of the pair,
$\mbox{\boldmath $\sigma$}_n$ and 
$\mbox{\boldmath $\tau  $}_n$.
Unless, the variational solution for the ground state only gives the state
with $b=0$, since the exact ground state is the left-dimer state on 
$\lambda$ bonds.
Then, the variation with $b=0$ gives nothing at all.
It should be noticed that 
a basic recipe of the present variational method is that a spin pair,
$\mbox{\boldmath $S$}_n = 
\mbox{\boldmath $\sigma$}_n + \mbox{\boldmath $\tau  $}_n$,
should be chosen so that it does not take a singlet dimer state.
Therefore, we shift $\mbox{\boldmath $\sigma$}$ spins by one site as
$\mbox{\boldmath $\sigma$}_n \to \mbox{\boldmath $\sigma$}_{n+1}$.
With this new definition of 
$\mbox{\boldmath $\sigma$}_n$ and $ \mbox{\boldmath $\tau  $}_n$,
the left-dimer ground state is represented by 
$b=\sqrt{3}/2$, or in other words, $a=0$.
Expressions of
$E_1$ and $E_2$ become different from Eqs.(\ref{eq:e1}) and (\ref{eq:e2}),
and are $E_1=-\lambda/2-1/4$ and $E_2=\lambda$, respectively.
The excitation is calculated dispersionless as 
\begin{equation}
 E_{\rm ex}(k)=\lambda.
 \label{eq:gaplocal2}
\end  {equation}
This result is nothing but the local triplet excitation,
where one singlet dimer is replaced by a triplet in the ground state.
This becomes the exact solution in the limit $\lambda \to \infty$.

\begin{figure}[h]
 \epsfxsize=8.0cm
 \epsffile{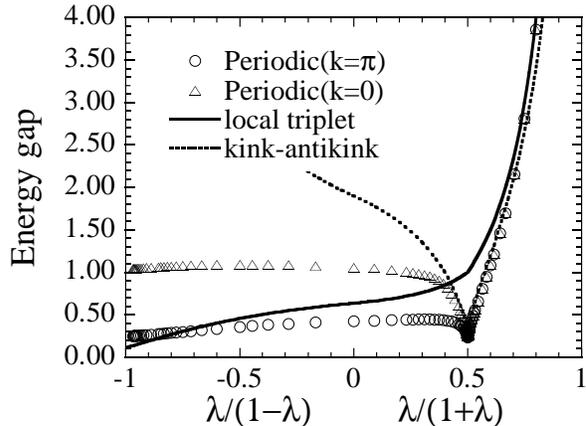}
 \caption{$\lambda$ dependence of the energy gap obtained by 
          the numerical diagonalization (circles),
          the variation of a kink-antikink type (broken line), and
          the variation of a local triplet type (solid  line).
          The second excitation gap with $k=0$  (triangles)
          is also plotted for $\lambda < 1$.
          $x$ axis is $\lambda /(1-\lambda)$ for $\lambda <0$, and 
                      $\lambda /(1+\lambda)$ for $\lambda >0$.
 \label{fig:gapall}
         }
\end  {figure}

Figure \ref{fig:gapall} shows the $\lambda$ dependence of the energy gap 
obtained by the variation and 
the numerical diagonalization of the periodic system with 24 spins.
The numerical results are depicted by symbols.
The variational results of a kink-antikink type excitation,
Eq. (\ref{eq:ec1}) and (\ref{eq:ec2}), are shown by a broken line;
those of a local triplet excitation, Eq. (\ref{eq:gaplocal1}) and
(\ref{eq:gaplocal2}), are shown by a solid line.
We also plot the second excitation gap for $\lambda \le 1$ for the
comparison with Eq. (\ref{eq:ec2}).

In the dimer phase, $\lambda > 1$,
consistency between the numerical and the variation is excellent.
Estimations by the kink-antikink variation is better 
in the vicinity of $\lambda =1$, while those of the local triplet 
becomes better as $\lambda$ increases.
This crossover occurs at $\lambda \sim 3$, where the average distance
between a kink and an antikink $\langle x\rangle$ of Eq. (\ref{eq:x1}) 
becomes unity.
On the other hand, the consistency in the region, $\lambda < 1$, only 
remains in a qualitative level.
Within the single-site approximation employed in this paper, 
$\mbox{\boldmath $\sigma$}$ spins and 
$\mbox{\boldmath $\tau  $}$ spins are equivalent to each other.
For example, 
$\langle n(b)|
\mbox{\boldmath $\sigma$}_n \cdot\mbox{\boldmath $\sigma$}_{n+1}
|n(b)\rangle$
is equal to
$\langle n(b)|
\mbox{\boldmath $\tau$}_n \cdot\mbox{\boldmath $\tau$}_{n+1}
|n(b)\rangle$.
Therefore, the approximation should be better for the system with 
a symmetry of exchanging 
$\mbox{\boldmath $\sigma$} \leftrightarrow \mbox{\boldmath $\tau  $}$.
Since the present model does not possess this symmetry, the estimation 
is not good.
It should be excellent in the symmetric systems such as
the Majumdar-Ghosh model.\cite{nakamura-ot96}
We must go beyond the single-site approximation to improve the 
estimates in the Haldane phase of the $\Delta$ chain.
The numerical data converges to the value, $\sim 0.24$ consistent 
with the $S=1$ system;
a half the gap of a typical diagonalization result of $S=1$ system 
with 12 spins, $0.4842$.
\cite{sakai-t90} 
Only the number of interaction bonds between $\mbox{\boldmath $S$}_n$
and $\mbox{\boldmath $S$}_{n+1}$ determines the strength of the 
effective interaction,
since
$\mbox{\boldmath $\sigma$}_n$ and $ \mbox{\boldmath $\tau  $}_n$
becomes symmetric in the $S=1$ limit.
The present model has only two interaction bonds, which corresponds to
a half the effective interaction in the $S=1$ system.

\section{Observables}
\label {sec:obs}

We calculated the magnetic susceptibility and the specific heat 
of finite system ($N=7$) with periodic boundary conditions 
in order to see the difference of observable physical quantities
between the dimer phase and the Haldane phase.
In other words, we try to determine the phase by these observables.
\begin{figure}[h]
 \epsfxsize=8.0cm
 \epsffile{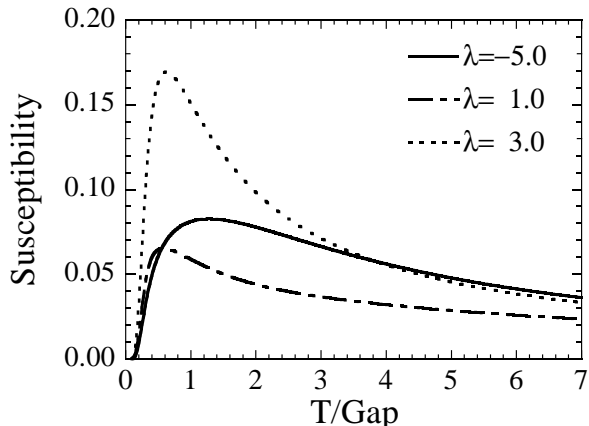}
 \caption{Uniform susceptibility 
          per spin calculated for $\lambda=3.0, 1.0$ and $-5.0$
          by the exact diagonalization of the finite systems with the 
          periodic boundary conditions.
          The size of the system is $N=7$.
          Each $\lambda$ represents the dimer phase, the 
          fully-frustrated point, and the Haldane phase, respectively.
 \label{fig:suscep}
         }
\end  {figure}

Figure \ref{fig:suscep} shows the uniform magnetic susceptibility
for $\lambda=-5.0, 1.00$ and $3.0$.
These parameters correspond to the Haldane phase, fully-frustrated point, 
and the dimer phase, respectively.
We rescale the temperature by each energy gap in order to see the 
qualitative differences, 
namely $E\to (E-E_{\rm g})/E_{\rm gap}$.
The peak width and the height among three are quite different.
Data of the dimer phase have a sharp peak, while it becomes broad in 
the Haldane phase.
This reflects the structure and the density of the excited states,
i.e., many continuum states of multiple-magnon excitations
in the Haldane phase bring about a broad peak.
Contrary,
excitations in the dimer phase are considered to be discrete 
local triplet excitations, which generates a rather narrow peak.
Full width at half maximum (FWHM) of the peak for each data 
in the unit of $T/{\rm Gap}$ is:
$\sim 2$ for $\lambda=3.0$,
$\sim 4$ for $\lambda=1.0$, and
$\sim 6$ for $\lambda=-5.0$.
The FWHM is almost three times as wide in the Haldane phase as in the 
dimer phase.
We speculate that
this value might be a judge to determine the phase. 
Recently, we have also calculated the susceptibility of the 
frustrated ladder model, and the $J_1$-$J_2$ model with bond dimerization
($J_1$-$J_2$-$J_3$ model), and found that the
FWHM take almost the same value as the present case.
Details are reported elsewhere.\cite{nakamura-ot96}

\begin{figure}[h]
 \epsfxsize=8.0cm
 \epsffile{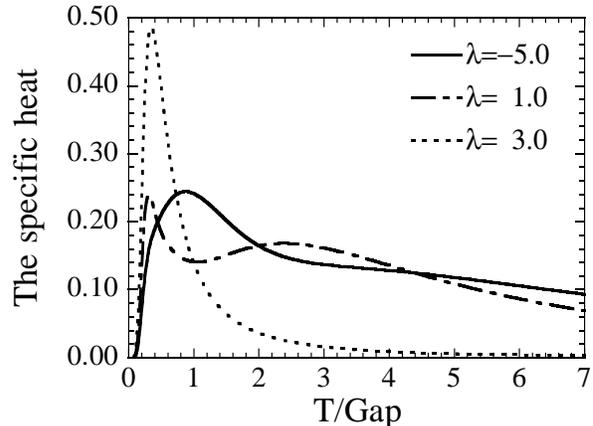}
 \caption{The specific heat
          per spin calculated for $\lambda= 3.0, 1.0$ and $-5.0$
          by the exact diagonalization of the finite systems with the 
          periodic boundary conditions.
          The size of the system is $N=7$.
 \label{fig:spec}
         }
\end  {figure}

Figure \ref{fig:spec} shows the specific heat.
Qualitative tendency is same as the susceptibility, except for
the data of $\lambda=1.00$ showing the double peak structure, which is
characteristic of the kink-antikink excitation.\cite{nakamura-k96}
Data of the dimer phase is explained by the Schottky type 
specific heat.

\section{summary and discussion}
\label{sec:sum}
We have investigated the $\Delta$ chain with the bond dimerization.
The model interpolates the independent dimer model and the $S=1$ spin 
chain model by changing $\lambda$.
As for the ground state, the first order transition between the dimer 
phase and the Haldane phase occurs at $\lambda=1$.
To be strict,
the ground state for $\lambda > 1$ is the perfect singlet dimer state, 
while that for $\lambda <1$ continuously changes from the dimer state 
at $\lambda =1$ to the $S=1$ Haldane state at $\lambda =-\infty$.

As for the excited state, we can distinguish the intermediate region,
in the vicinity of $\lambda =1$, from the dimer region and the Haldane region.
This region is characterized by strong frustration.
The energy gap is very small caused by the unstable ground state.
The ground state has a strong dimer correlation.
The excited state is governed by two free spin called 
a kink and an antikink.
They become a local triplet in the dimer and the Haldane region.
We can conclude that the intermediate region serves as a buffer
between the other two regions.

We clarified the wave function of the bound state of a kink and an antikink,
and how they collapse to a local triplet.
The essential point is that a kink and an antikink is bound by a 
triangular potential and the competition between this attractive 
potential and the kinetic energy causes the gap enhancement with 
exponent $2/3$.
The triangular potential is an outcome of unfavored singlet dimers
between a kink and an antikink.
This mechanism generally occurs when the frustration is relaxed 
by bond dimerization.
For example,
Chitra et al investigated the 1-dimensional $J_1$-$J_2$ model with 
the bond alternation, and found that the gap behaves with $\delta^{2/3}$ with 
the alternation parameter $\delta$.\cite{chitra95}
This can be explained in the same manner as in the present model.

We have also calculated the uniform susceptibility and the specific heat 
to see how the phases are characterized by the physical observables.
We clarified the qualitative difference between the Haldane phase 
and the dimer phase.
The susceptibility data shows a broad peak in the Haldane phase, 
accompanied by continuum multiple magnon excitations.
Therefore the FWHM of the peak may serve to judge which phase the data
belong to.

\acknowledgments

Authors 
would like to thank Professor K. Kubo and Professor 
A. Oshiyama for valuable discussions.
They also 
acknowledge thanks to Professor H. Nishimori for his 
diagonalization package, Titpack Ver. 2.
Computations were performed partly 
on Facom VPP500 at the ISSP, University of Tokyo.

\appendix
\section{Non-local unitary transformation}
\label{app:nlu}
In this appendix, we summarize the non-local unitary transformation for
the double $S=1/2$ spin-chain systems.
The transformation is defined by $U$ in the following.
\begin{eqnarray}
 U&=&\prod_{n=1}^{N} U_n, \\
 U_n&=&P_n^+ + P_n^- \exp[i\pi S_n^x],     \\
 P_n^{\pm}&=&\frac{1}{2}\left(1\pm \exp
                       \left[i\pi\sum_{k=1}^{n-1}S_k^z\right]\right),\\
 \mbox{\boldmath $S$}_n &=& \mbox{\boldmath $\sigma$}_n 
                        + \mbox{\boldmath $\tau  $}_n,
\end  {eqnarray}
where $P_n^+$ ($P_n^-$) is the projection operator onto
states with the even (odd) number of $S_i^z=\pm 1$ for $i\le n-1$.
The spin operators, $\sigma_n^x$, $\sigma_n^y$, $\sigma_n^z$, are 
transformed as,
\begin{eqnarray}
 U\sigma_n^x U &=& \exp \left[ i\pi\sum_{m=n+1}^N   S_m^x\right]\sigma_n^x, \\
 U\sigma_n^y U &=& \exp \left[ i\pi\sum_{k=1}^{n-1} S_k^z\right]
                   \exp \left[ i\pi\sum_{m=n+1}^{N} S_m^x\right]\sigma_n^y, \\
 U\sigma_n^z U &=& \exp \left[ i\pi\sum_{k=1}^{n-1} S_k^z\right]\sigma_n^z.
\end  {eqnarray}
In the above derivation, the following relations are utilized:
\begin{eqnarray}
  U^{-1} = U,  \nonumber \\
  U_n U_m= U_m U_n,              \nonumber \\
  (\exp[i\pi S^{\alpha}])^2=1,  \nonumber \\
  \sigma_n^xP_n^{\pm} = P_n^{\mp}\sigma_n^x,  \nonumber \\
  \sigma_n^{\alpha}\exp[i\pi S_n^{\beta}] =
 -\exp[i\pi S_n^{\beta}] \sigma_n^{\alpha} \mbox{ ($\alpha \ne \beta$)}, 
\nonumber \\
  \exp[i\pi S_n^{\alpha}]
\exp[i\pi S_n^{\beta }]
\exp[i\pi S_n^{\gamma}]=1 \mbox{ (for cyclic $\alpha, \beta, \gamma$)}, 
\nonumber
\end  {eqnarray}
which are easily obtained by using
$\exp[i\pi \sigma_n^{\alpha}]=2i\sigma_n^{\alpha}$.
Transformations for $\mbox{\boldmath $\tau$}_n$ are obtained by replacing 
$\sigma$ by $\tau$.
By using the above relations, exchange interactions becomes,
\begin{eqnarray}
 U\mbox{\boldmath $\sigma$}_n \cdot \mbox{\boldmath $\tau  $}_n U &=&
  \mbox{\boldmath $\sigma$}_n \cdot \mbox{\boldmath $\tau  $}_n, \nonumber \\
 U\mbox{\boldmath $\sigma$}_n \cdot \mbox{\boldmath $\sigma$}_{n+1} U &=&
   - \sigma_n^x\tau_{n+1}^x - \tau_n^z\sigma_{n+1}^z
   -4\sigma_n^x\tau_{n+1}^z   \tau_n^z\sigma_{n+1}^z,        \nonumber     \\
 U\mbox{\boldmath $\tau  $}_n \cdot \mbox{\boldmath $\sigma$}_{n+1} U &=&
   - \tau  _n^x\tau_{n+1}^x - \sigma_n^z\sigma_{n+1}^z
   -4\tau  _n^x\tau_{n+1}^x   \sigma_n^z\sigma_{n+1}^z.     \nonumber
\end  {eqnarray}
\section{Second-order perturbation of the ground state energy 
         for $\lambda < 1$}
\label{app:2ndp}

In this appendix, we show that the second-order perturbation of the
ground state energy can be done in this case without knowing an excited
state.
We divide the Hamiltonian ${\cal H}$ into the unperturbed part ${\cal H}_0$ 
and the perturbation part ${\cal H}_1$:
$ {\cal H}={\cal H}_0 + {\cal H}_1$,
where 
\begin{eqnarray}
  {\cal H}_0&=&\sum_{n=1}^N  
     \mbox{\boldmath $\sigma$}_n \cdot \mbox{\boldmath $\tau  $}_n
    +\mbox{\boldmath $\sigma$}_n \cdot \mbox{\boldmath $\sigma$}_{n+1}
    +\mbox{\boldmath $\tau  $}_n \cdot \mbox{\boldmath $\sigma$}_{n+1},\\
  {\cal H}_1&=&(\lambda-1)\sum_{n=1}^{N-1}
     \mbox{\boldmath $\tau  $}_n \cdot \mbox{\boldmath $\sigma$}_{n+1}.
\end  {eqnarray}
Here, note that the location of $\mbox{\boldmath $\sigma$}_n$ and 
$\mbox{\boldmath $\tau  $}_n$ is different from Eq. (\ref{eq:transhamil})
and Fig. \ref{fig:lattice} 
so that 
the pair $\mbox{\boldmath $S$}_n =  
          \mbox{\boldmath $\sigma$}_n + \mbox{\boldmath $\tau  $}_n$
takes the singlet dimer state for the unperturbed Hamiltonian.
The unperturbed ground state $|\psi_0\rangle$ is a direct product of 
singlet dimer states $|S_n\rangle$ on the spin pair 
$\mbox{\boldmath $\sigma$}_n$ and $\mbox{\boldmath $\tau  $}_n$, i.e.,
\begin{equation}
 |\psi^{0}\rangle  =  \prod_{n=1}^N |S_n\rangle.
\end  {equation}

Let us first examine the first order perturbation $E_1$:
\begin{equation}
  E_1=\langle\psi^{0}|{\cal H}_1|\psi^{0}\rangle = 
     (\lambda -1)\sum_{n=1}^{N-1}\langle \psi^0|\psi ^1 _n\rangle,
  \label{eq:app-7}
\end{equation}
where
\begin{eqnarray}
  |\psi^1 _n\rangle &=&
     \mbox{\boldmath $\tau  $}_n \cdot \mbox{\boldmath $\sigma$}_{n+1}
  |\psi^0\rangle  \nonumber \\
  &=& -\frac{\sqrt{3}}{4}\prod_{k\ne n, n+1}^{N}
  |S_k\rangle |T^2_n\rangle .
  \label{eq:app-8}
\end  {eqnarray}
$|T^2_n\rangle$ denotes a singlet state formed by two triplet pairs of
$\mbox{\boldmath $S$}_n$ and $\mbox{\boldmath $S$}_{n+1}$, 
i.e.,
\begin{equation}
 |T^2_n\rangle = (
  |  0 _n\rangle |  0 _{n+1}\rangle  
 -|  1 _n\rangle |(-1)_{n+1}\rangle 
 -|(-1)_n\rangle |  1 _{n+1}\rangle )/\sqrt{3}.
\end  {equation}
$|0_n\rangle, |1_n\rangle$ and $|(-1)_n\rangle $ denote 
triplet states of a pair 
$\mbox{\boldmath $S$}_n = 
\mbox{\boldmath $\sigma$}_n + \mbox{\boldmath $\tau  $}_n$ with 
their eigenvalues of  $ S_n^z=\sigma_n^z + \tau_n ^z$.

It follows from Eqs. (\ref{eq:app-7}) and (\ref{eq:app-8}) that 
\begin{equation}
 E_1=0,
\end  {equation}
since the singlet state $|S_n\rangle$ is orthogonal to the triplet state.
It is found in Eq. (\ref{eq:app-8}) that the operation 
$\mbox{\boldmath $\tau  $} _n \cdot \mbox{\boldmath $\sigma$}_{n+1}$
on $|\psi ^0\rangle$ transforms the two singlets $|S_n\rangle $ and 
$|S_{n+1}\rangle $ into triplets leaving the other singlets unchanged.

The second-order perturbation $E_2$ is calculated as
\begin{eqnarray}
 E_2&=&-\langle\psi^0|{\cal H}_1 \frac{1}{{\cal H}_0-E_0}{\cal H}_1
              |\psi^0\rangle  \nonumber \\
    &=&-(\lambda-1)^2\sum_{n,m=1}^{N-1} 
     \langle\psi^1_m|\frac{1}{{\cal H}_0-E_0}|\psi^1_n\rangle  \nonumber \\
    &=&\frac{(\lambda-1)^2}{E_0}\sum_{n,m=1}^{N-1}\sum_{k=0}^{\infty}
     \langle\psi^1_m|\left (\frac{{\cal H}_0}{E_0}\right )^k|\psi^1_n\rangle .
\end  {eqnarray}
To proceed further, we divide ${\cal H}_0$ into a diagonal part 
and an off-diagonal part when it operates to $|\psi^1_n\rangle $.
An off-diagonal part ${\cal H}_n^{\rm OD}$ is
\begin{equation}
 {\cal H}^{\rm OD}_{n}= (
  \mbox{\boldmath $\sigma$}_{n+1} + \mbox{\boldmath $\tau  $}_{n+1} )
  \cdot
  \mbox{\boldmath $\sigma$}_{n+2}.
\end  {equation}
A diagonal part is the rest of the Hamiltonian, 
${\cal H}_0-{\cal H}^{\rm OD}_{n}$.
The eigenvalue of this diagonal part is $E^{\rm D}_1=E_0+1=-3N/4+1$.
On the other hand, the operation of 
${\cal H}_n^{\rm OD}$ to $|\psi_1\rangle _n$
generates a state with three triplets located at the $n$th, the $(n+1)$th, 
and the $(n+2)$th triangles, and these triplets form a singlet state.
Namely,
\begin{equation}
 |\psi^2_n\rangle ={\cal H}^{\rm OD}_{n}|\psi^1_n\rangle 
                 =\frac{1}{\sqrt{2}}\prod_{k\ne n, n+1, n+2}
                  |S_k\rangle  |T^3_n\rangle ,
\end  {equation}
where the three-triplets state
\begin{eqnarray}
 |T^3_n\rangle &=&\frac{1}{\sqrt{6}}[
(| 1_{n+2}\rangle |(-1)_{n+1}\rangle -|(-1)_{n+2}\rangle | 1_{n+1}\rangle )
 \times             | 0_{n  }\rangle   
  \nonumber \\
&+&
(| 1_{n+1}\rangle |(-1)_{n  }\rangle -|(-1)_{n+1}\rangle | 1_{n  }\rangle )
  \times             | 0_{n+2}\rangle   
  \nonumber \\
&+&
(| 1_{n  }\rangle |(-1)_{n+2}\rangle -|(-1)_{n  }\rangle | 1_{n+2}\rangle )
  \times             | 0_{n+1}\rangle   
  ]
\end  {eqnarray}
forms a singlet state. 
Now we get 
\begin{equation}
 {\cal H}_0|\psi^1_n\rangle  = E_1^{\rm D}|\psi^1_n\rangle + |\psi^2_n\rangle.
\end  {equation}
Similarly, we find
\begin{equation}
 {\cal H}_0|\psi^2_n\rangle  = E_2^{\rm D}|\psi^2_n\rangle + |\psi^3_n\rangle,
\end  {equation}
where $E_2^{\rm D}$ denotes diagonal energy and the $|\psi^3_n\rangle $
is a new state with four triplets forming a singlet state located at the
triangle site from $n$ to $n+3$.
In general, we have
\begin{equation}
  ({\cal H}_0)^k|\psi^1_n\rangle  
  = (E_1^{\rm D})^k|\psi^1_n\rangle  + \sum_{l=2}^{k+1}C_l|\psi^l_n\rangle,
\end  {equation}
where $|\psi^l_n\rangle $ contains $l+1$ triplets at triangle sites 
$n \sim n+l$ forming a singlet state; and
$C_l$ is a constant.
Thus we obtain
\begin{eqnarray}
 \langle\psi^1_m|{\cal H}_0^k|\psi^1_n\rangle  &=&
  \delta_{n,m}(E_1^{\rm D})^k \langle\psi^1_m|\psi^1_n\rangle  \nonumber \\
  &=&
  \delta_{n,m}(E_1^{\rm D})^k \frac{3}{16}.
\end  {eqnarray}
Finally, we get the energy correction of the second order
\begin{eqnarray}
 E_2&=&\frac{3}{16}\frac{(\lambda-1)^2}{E_0}\sum_{n=1}^{N-1}\sum_{k=0}^{\infty}
     \left (\frac{E_1^{\rm D}}{E_0}\right )^k \nonumber \\
    &=&-\frac{3}{16}(\lambda -1)^2(N-1)\frac{1}{E_1^{\rm D}-E_0} \nonumber \\
    &=&-\frac{3}{16}(\lambda -1)^2(N-1).
\end  {eqnarray}
This agrees with the variational result given by Eq. (\ref{eq:vareg}) 
in the text.

\end{multicols}
\end{document}